\documentclass[aip,rsi,reprint,amsmath,amssymb,floatfix]{revtex4-1}
\usepackage{graphicx}  
\usepackage{dcolumn}   
\usepackage{bm}        
\usepackage{amssymb}
\usepackage{amsmath}
\usepackage{tabu}
\usepackage{epsfig}
\usepackage{epstopdf}
\setcitestyle{super}
\usepackage{enumerate}

\begin{document}

\title{A Scanning, All-Fiber Sagnac Interferometer for High Resolution Magneto-Optic Measurements at 820 nm}

\author{Alexander Fried}
\affiliation{Department of Physics, Stanford University, Stanford, CA 94305}
\affiliation{Geballe Laboratory for Advanced Materials, Stanford University, Stanford, CA 94305}
\author{Martin Fejer}
\affiliation{Department of Applied Physics, Stanford University, Stanford, CA 94305} 
\affiliation{Ginzton Laboratory, Stanford University, Stanford, CA 94305}
\author{Aharon Kapitulnik}
\affiliation{Department of Applied Physics, Stanford University, Stanford, CA 94305} 
\affiliation{Department of Physics, Stanford University, Stanford, CA 94305}
\affiliation{Geballe Laboratory for Advanced Materials, Stanford University, Stanford, CA 94305}

\date{\today}

\begin{abstract}
The Sagnac Interferometer has historically been used for detecting non-reciprocal phenomena, such as rotation.  We demonstrate an apparatus in which this technique is employed for high resolution measurements of the Magneto-Optical Polar Kerr effect\textemdash a direct indicator of magnetism.  Previous designs have incorporated free-space components which are bulky and difficult to align.  We improve upon this technique by using all fiber-optic coupled components and demonstrate operation at a new wavelength, 820 nm, with which we can achieve better than 1 $\mu$rad resolution.  Mounting the system on a piezo-electric scanner allows us to acquire diffraction limited images with 1.5 $\mu$m spatial resolution.  We also provide extensive discussion on the details and of the Sagnac Interferometer's construction.

\end{abstract}

\maketitle

\section{Introduction}

Magneto-optical (MO) effects are described within quantum theory as the interaction of photons with electrons through the spin-orbit interaction\cite{pershan}.  Macroscopically, linearly polarized light that interacts with magnetized media can exhibit both ellipticity and a rotation of the polarization state. The leading terms in any MO effect are proportional to the antisymmetric off-diagonal part of the ac conductivity: $\sigma_{xy}(\omega) = \sigma_{xy}^\prime (\omega) + i \sigma_{xy}^{\prime \prime}(\omega)$ \cite{argyres}.  If a magnetic field is applied to the material, $\sigma_{xy}(\omega,\textbf{H})$ is small and proportional to the field. Its zero frequency limit is the known Hall conductivity of the material. If the material exhibits finite magnetization, $\sigma_{xy}(\omega,\textbf{M})$ is also small and proportional to the magnetization, and its zero frequency limit is known as an example of the anomalous Hall effect. Other, less straightforward effects of magnetism in a material will also lead to finite MO effect. Thus, a finite MO effect measured in a material points to time-reversal symmetry breaking (TRSB) in that system. 

Magnetization normal to the surface of a reflective sample can be measured optically via the Magneto-Optical Polar Kerr Effect.  Specifically, polarization rotation is characterized by the Kerr angle, which, in the circular basis, is half the difference in the phase accumulated, upon normal incidence reflection, between right and left circularly polarized light.  Provided non-linear optical effects are weak, the Kerr angle is proportional to the magnetization.  Because optical interactions are strongly dependent on electronic band structure and spin-orbit coupling parameters, the proportionality constant varies with temperature and optical frequency.  As an example, optical resonances, such as the Fermi-Edge singularity in GaAs, can enhance or even switch the sign of the Kerr angle even if the magnetization is constant.\cite{crooker,palik,mahan}

While the widely used cross-polarization techniques offer some advantages at measuring Kerr rotation\cite{drew,goa}, the past couple decades have seen their angular resolution limits bested by new strategies, as demonstrated by cavity-enhanced Kerr rotation for quantum dots,\cite{mete} ``optical bridge photodetection," \cite{kato,kato1} and, in particular, a modified Sagnac Interferometer (SI), first introduced by Spielman {\it et al.} \cite{spielman1}, in order to optically search for anyons in high-temperature superconductors\cite{anyons}.  While many of the above techniques can achieve Kerr angle resolutions better than 1$\mu$rad/$\sqrt{Hz}$, only the SI technique allows for a direct high-resolution detection of the MO effect without any external modulation of an external field that couples to the measured magnetic signal such as magnetic field, electric field, or current. This property is of utmost importance when intrinsic TRSB effects need to be detected, such as in search for such effects in unconventional superconductors.\cite{kapitulnik1}

\section{The Sagnac Interferometer}

Complicating any accurate measurement of the Kerr angle are non-magnetic effects, such as linear birefringence and linear dichroism, which can also affect the polarization state of light.  These effects can arise not only in the sample, but in the apparatus itself and prudence requires consideration of their existence.  A Sagnac interferometer, however, is sensitive only to non-reciprocal effects\cite{potton,shelankov, lefevre} as we will discuss in detail below, which in materials can arise from TRSB phenomena, such as magnetism\cite{dodge1}.  With this as a guiding principle, given that the optical components are reciprocal even if they are misaligned or imperfect, any signals measured will be strictly indicative of non-reciprocity arising from the material being examined.  Nonlinear optical effects are an exception and they can cause nonreciprocal polarization rotation as they are outside of the scope of reciprocity arguments that apply to only linear processes\cite{Trzeciecki}.  However, as nonlinear effects are expected to have an intensity dependence, they can easily be identified.  We discuss this in more detail within Appendix \ref{nonlinearAppendix}. 

\begin{figure} [h]
\includegraphics[width=1.0 \columnwidth]{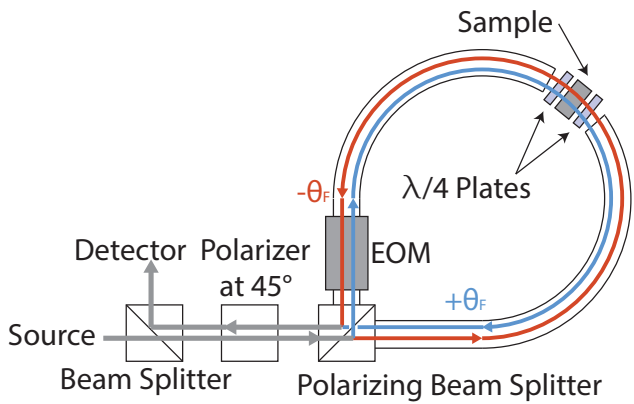}
\caption{Simplified illustration of the original Sagnac interferometer in the Faraday geometry.  The first polarizer orients the source polarization so that the beam splitter evenly couples the light into the two time reversed equivalent and therefore, reciprocal, illustrated by the inner (blue) and outer (red) channels in the loop.  The signals interfere destructively at the beam splitter after a single traversal of the loop if the sample is non-reciprocal and the two beams yield different phase shifts.  The modulated intensity is measured by lock-in detection and the amplitudes of the harmonics are used to calculate the Faraday angle, $\theta_F$.  This design can also be adapted to the normal and oblique incidence Kerr geometries\cite{spielman2}.}
\label{sagnac2}
\end{figure}

In a Sagnac interferometer, such as the one used by Dodge {\it et al.}\cite{dodge1} or  Spielman {\it et al.}\cite{spielman2} polarized light from a source at wavelength $\lambda$ is passed through a polarizer and launched into the slow axis of a single-mode, polarization-maintaining (PM) optical fiber.  The light is then split into two beams by a loop coupler (a fiber beam-splitter). These two beams propagate in opposite directions around the loop, which contains the sample as shown in Figure \ref{sagnac2}.  In general, if the loop is a reciprocal path (i.e. if its optical path length does not depend upon the direction of propagation), then the two beams will return to the loop coupler precisely in phase and constructively interfere at the detector.  If elements within the loop are nonreciprocal, the two beams yield a phase difference.  If the phase shifts originate from a Faraday effect, this phase difference will appear as destructive interference which will reduce the detected intensity to a fraction $[1+\text{cos}(2\theta_F)]/2$ of the maximum. 

In this paper we use the convention of defining circular polarization in terms of spin angular momentum and not helicity.  Thus, a circular polarization state of the electric field is defined as positive or negative by $\pm \sim \hat{\mathbf{x}}\pm i \hat{\mathbf{y}}$ with respect to a fixed coordinate axis.  Light passing through the sample must be in circular polarization states; opposite circular polarizations, traveling in opposite directions are a pair of time-reversed states of light, and will be affected differently by the sample only if the sample breaks TRS, which in materials, occurs if magnetism is present.  As the light emerging from the fiber is in a linear polarization state, a quarter-wave plate, with slow axis oriented at $45^\circ$ with respect to the fiber's slow axis can be aligned so that it converts the linearly polarized light in the clockwise (counter-clockwise) mode into positive (negative) circularly polarized light when it emerges from the fiber, and the second waveplate re-converts the beams back to linear polarization states, but now rotated $90^\circ$.  Since the only light on the slow axis of the fiber and the two beams must be in opposite circular polarizations at the sample, the outputs of the two fibers must be aligned so that their slow axes are also $90^\circ$ from each other.  Any remaining light that couples to the fast axis of the fiber is filtered out by the polarizer.

To enable the measurement of $2\theta_F$ to $\mu$rad sensitivity, the system is actively biased by an electrooptic phase modulator (EOM) in the loop, which sinusoidally modulates the index of refraction of the nonlinear crystal within it.  The modulator is operated at a frequency $f$ with period corresponding to twice the optical transit time of the full length of the loop.  The time-varying phase shifts induced on the two beams by the modulator are equal and opposite, since they pass through the modulator at times separated by half the modulator period.  From Equation \ref{intensity}, to first order in $2\theta_F$, the signal from the detector will contain a second harmonic of $f$, with amplitude proportional to the throughput of the loop ($\mathcal{I}$), and a first harmonic with amplitude proportional to $2\theta_F\times \mathcal{I}$, the quantity of interest.  These harmonics are measured with lock-in amplifiers and divided to compute the Faraday angle. 

The Reciprocity Principle states that the results of a measurement are identical to another performed under the conditions of time being reversed and the positions of the optical source and the optical detector interchanged.  There are various uses of this terminology in the literature.  A ``reciprocal material" is one which acts on light in the same fashion if only the source and detector positions are exchanged.  A pair of ``reciprocal optical paths" refer to the polarization state and path of light taken between a light source and detector, and the polarization state and path that would be traversed if time was reversed (i.e. if the the source and detector were exchanged and the conjugate operation performed on the complex electromagnetic field).  

\begin{figure} [h]
\includegraphics[width=1.0 \columnwidth]{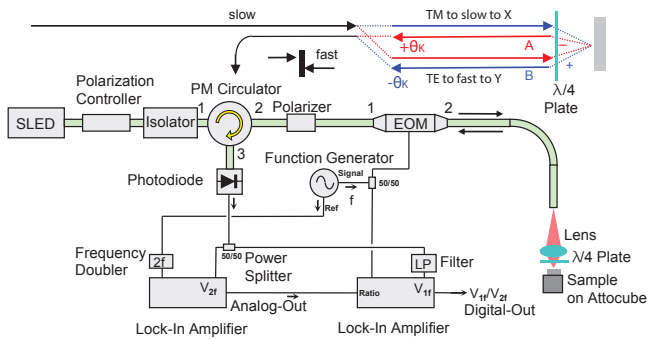}
\caption{Schematic of the interferometer.  The ``A" (red) and ``B" (blue) paths commensurate with the fiber components illustrate the two counter-propagating beams in the zero-area loop that forms the interferometer.  The counter-propagating beams illustrated in the upper part of the Sagnac loop are traversing along the TM mode of the EOM and coupling to and from the slow axis of the 10m PM fiber and to the X-polarization state in free space.  The lower part of the loop carries  describes light along the TE mode of the EOM which couples to the fast axis of the fiber and exits into the Y-polarization state.}
\label{sagnac}
\end{figure}

A Faraday rotator is a nonreciprocal device because it is magnetic and breaks time reversal symmetry, whereas a quarter-wave plate is reciprocal, since its index of refraction is only dependent on the linear polarization state.  Likewise, the two beam paths shown in Figures\ref{sagnac2} and \ref{sagnac} which make up the Sagnac Interferometer are reciprocal because they are identical if the time reversal operation is applied to one of them.  Any differences between the phases or amplitudes of these two beams, when measured at the detector after traversing the loop, are signatures of non-reciprocity.  The Sagnac interferometer is sensitive only to the phase differences, and since there are nominally no non-reciprocal effects in the fiber and the other optical components, the only source of non-reciprocity will be at the sample.

An improvement in design and performance of the basic Sagnac interferometer was implemented by Xia {\it et al.},\cite{xia1} by using both the fast and slow axes of a PM fiber as optical paths, thus forming an SI with a zero-area-loop.  Furthermore, this setup applied the technique to a measurement in the Kerr geometry.  To introduce active biasing, free space optics direct light from the source through a polarization dependent phase modulator, before proceeding to the fiber, a single quarter-wave plate and the sample.  Because the loop area is zero, non-reciprocal effects stemming from any mechanical rotation, including the Earth's rotation, will not appear.

We have further extended upon the work of Xia {\it et al.}  by using a new all-fiber approach which dramatically simplifies the alignment, has a small footprint and can be constructed quickly from off-the-shelf components.  Like its predecessor, this design allows for flexible temperature capability and can be ported into a wide variety of experimental designs.  Our source is a superluminescent diode (SLED) centered at 818 nm, temperature controlled at $25^\circ$C, and possessing a bandwidth of about 22 nm.  Using both Michelson interferometry and spectral analysis, we measure the coherence length to be about 8 $\mu$m over a wide range of output power.  We operate the SI with about 180 $\mu$W of power incident on the sample, with only about 20$\mu$W returning to the detector when the sample is a gold mirror.  The low throughput to the detector, following reflection from the sample, is a consequence of inefficient coupling and focusing of light from the fiber to the sample.  A custom lens would offer better performance, as would, larger, off-the-shelf fiber-coupling components which are easier to align.

As shown in Figure~\ref{sagnac} the light from the superluminescent diode passes through a polarization controller, two PM isolators and a PM circulator, all optimally functioning for wavelengths centered at 820 nm.  Back-reflections to the SLED are the single most significant source of slow drifts in our measurements, so the two isolators and circulator provide 85dB of isolation from such effects.  At this level of isolation, no back-scatter is measured, nor does the addition of a third isolator gives makes no noticeable difference.  Light emerging from the second port of the circulator, will have power primarily on the slow axis of the fiber and any remaining light on fast axis is filtered away with the PM fiber polarizer.  

The critical custom design specifications for the EOM are (1) on Port 1, the slow axis of the fiber is oriented at $45^{\circ}$ to the TM mode of the its Lithium Niobate waveguide modulator and (2) the slow axis of the fiber is aligned to the high-index, TM, mode of the modulator on Port 2.  When the beam then proceeds to the custom EOM, the first requirement ensures light couples roughly equally to the TM and TE modes.  While the EOM modulates both modes, the modulation depth of the former is larger and the ``B" mode gains a net phase.  The second requirement helps suppress parasitic interferometric paths and is discussed in detail in the next section.

The TM and the TE modes in the EOM couple to the slow and fast modes, of a 10.01m long fiber and are then focused onto the sample by an aspherical lens with anti-reflection coatings.  The lens' numerical aperture on one side is matched to the fiber (0.12) and is 0.43 on the sample side.  An anti-reflection coated quarter-wave plate, with its slow axis at $45^\circ$ to the slow axis of the fiber, transforms the orthogonal, linear polarized modes emerging from the fiber to opposing circularly polarized modes.  When reflecting off a magnetic sample, these modes will yield different phases; half their difference will be the measured Kerr angle.

After reflecting, the second pass through the quarter-wave plate returns the modes to linearly polarized states, but with their polarizations now interchanged from before.  Consequently, the axes in the fiber into which they couple will also be exchanged, as shown in Figure \ref{sagnac}.  The beams return to the EOM where the ``A" beam will receive the dominant phase modulation.  When both beams re-couple to the slow axis of the fiber, they will interfere.  Any remaining power on the fast-axis is, again, filtered by the polarizer and the beam returns to the circulator and sends the light to a 125 MHz photodetector.  

We use two lock-in amplifiers to measure the first and second harmonics at 4.867 MHz and 9.734 MHz respectively and a low-pass filter is placed on the input of the former to remove the large second harmonics. The lock-in amplifier integration time restricts the time resolution; we typically set it to about 1 second.  The in-phase first harmonic signal is normalized by the second harmonic magnitude and the recorded output voltage of the lock-in is the quotient of the two.  This quotient is used to compute the Kerr angle, while the second harmonics are usually interpreted as the reflectivity.  However, it really corresponds to the power present in the two interfering modes.  There are many scattering processes such as imperfections in the equipment or birefringence of the sample which can scatter light out of these modes and decrease the second harmonic magnitude.  We review the details of this calculation and the specific alignment techniques in the Appendix \ref{calculations}.

\section{Competing Interferometers}

Essential to the technique reported in Xia et al. is the free-space electro-optical modulator (EOM).  This type of EOM introduces Residual Amplitude Modulation (RAM) in the form of beam steering, back-scattering, and etalon effects \cite{garzaella,meas} from piezoelectric resonances.  We observe that these generate drifting, spurious signals of several $\mu$rad on top of the desired signal and which depend exactly on the trajectory of the light beam through the EOM.  Nominally, because of the $\pi$ phase difference on the EOM drive signal for the forward and return paths, RAM should cancel itself out to first order and spurious signals should be reduced.  In a free-space EOM, it is difficult to perform an alignment that ensures that the forward and reverse paths not identical.  Another difficulty is that the resonant modulators used in previous versions require the fiber to be cut precisely to the length that corresponds to the resonant drive frequency.  

In an all-PM-fiber system there are only four propagating modes in both the fiber and the EOM.  At a given frequency, there are two polarizations traveling in two directions, so no beam-steering can occur.  Since the modulator is non-resonant, the fiber length does not matter as the drive frequency can always be adjusted without sacrificing the performance of the EOM.  Furthermore, if not terminated with 50 ohms, the inline EOM can operate at hundreds of MHz, allowing the fiber length to be shortened\cite{eospace}.  A detailed analysis of RAM is presented in Appendix \ref{RAMap}.

The two fiber components used in lieu of free-space optics are a PM fiber-coupled-polarizer made by Oz-Optics and a custom non-resonant EOM made by EOSPACE Inc.  The polarizer has extinction ratio 1:100 and is polarized along the slow axis of the fiber.  This EOM has large insertion loss; throughput to the sample is about 45$\%$ as compared with 70$\%$ or more with the free-space optics.  EOSPACE guarantees that they can align the fiber axes to waveguide modulator axes with an accuracy of less than 3$\%$, which will unevenly distribute light between the two modes in this component.  By collimating the output of the EOM and passing it through a half-wave plate on a rotation stage, a fixed 1:100000 extinction ratio Glam-Thompson polarizer, and finally, to a photodetector, we measured the power difference on both channels and find it to be less than 1$\%$.  The modulators also have power limits of around 20 mW, above which throughput decreases and performance will deviate from specifications.\cite{eospace}  The EOM generates RAM and transmission measurements reveal it to be sharply peaked at the driving frequency, with almost no spectral weight above it and some below it.  As in the free-space design, all fiber connectors are angle polished at $8^\circ$ to minimize back-reflections. 

\begin{figure} [h]
\includegraphics[width=.9 \columnwidth]{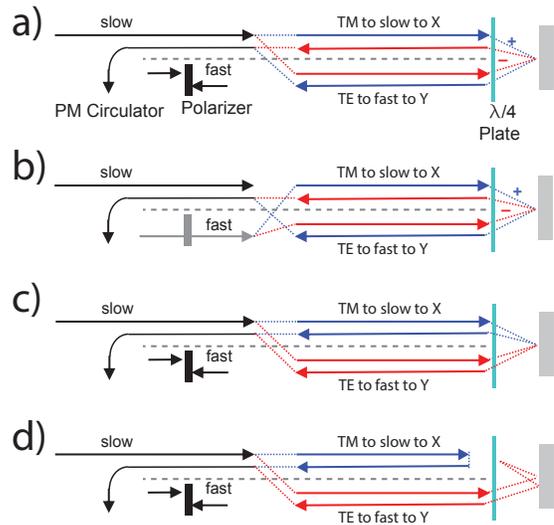}
\caption{A comparison of possible optical paths in the interferometer. (a) shows the two primary interfering paths also shown in Figure \ref{sagnac}.  If the polarizer or the circulator is imperfect, then three additional pairs of reciprocal paths can exist, which will also interfere with the primary two.  Depicted in (b) is one which shows light from the fast axes of the circulator, coupling to the EOM and returning to the slow axis of the fiber.  Light can also couple through the system starting in the slow axis and ending in the fast axis, as well as starting on the fast axis and ending back on the fast axis. Those paths illustrated in (c) will be incoherent with those in (a) and not interfere.  (d) There are points in the setup, where back-scattering will form optical paths with the same path-length as in (a), and thus can interfere with the primary signal.  However as these points are not near scattering facets, such scattering should not be present.}
\label{Paths}
\end{figure}

Because of imperfections and misalignments, there will always be modes of light which do not follow the ideal optical trajectories illustrated in Figure \ref{sagnac} and Figure \ref{Paths}a.  Several possible mechanisms are known to result in light traveling along non-ideal paths.  1) An imperfect polarizer will route some light along the fast axis of the EOM and, upon return, can couple light back into the fast or slow axes of the circulator as illustrated in Figure \ref{Paths}b.   2) Following the EOM, cross-coupling can appear either through scattering or fiber connectors that are imperfectly aligned.  3) The quarter-wave plate can be angularly misaligned, be of the wrong thickness, or in general, not functioning ideally as the incident light is not a plane wave.  4) The sample could be linearly birefringent, so the polarization eigenstates of reflection may not be circular.  The last three result in trajectories illustrated in Figure \ref{Paths}c.  5) Back-scattering at interfaces can reflect light backwards leading to competing interferometers.  As discussed above, angle polished components prevent back scattering, and no evidence of it has been observed, as there is no signal at the detector when the sample is removed.  6) The alignment of the slow axis of the polarizer might not be perfectly aligned at $45^\circ$ to the TM and TE modes of the EOM at port 1 and unequal power will be distributed along the two modes.  As the SI measures phase differences, differences in power along the two modes are unlikely to cause an effect, provided there are no nonlinear interactions.  In the same way, circular dichroism in the sample will also not contribute as it only affects the amplitude.  As through out the entire apparatus the PM fiber ensures that there are only two polarization channels are allowed in a given propagation direction, a detailed analysis of how any imperfection within the interferometer is accurately accomplished by using the Jones Calculus, as described in Appendix \ref{calculations}.

Two powerful principles can be drawn upon to quickly provide accurate qualitative understanding of how phases accumulate on the various optical pathways.  The first principle, is that, by definition, only non-reciprocal phenomena can cause pairs of otherwise reciprocal optical paths to yield phase differences upon a complete traversal of the instrument.  The fiber, the polarizer, the EOM, the wave plate and the rest of the optical components within the loop are non-magnetic, so only the sample is expected to be a source of non-reciprocity, and any deviation from the baseline Kerr signal is therefore expected to be a positive identification of non-reciprocity in the sample.  Reciprocity can also be used to understand the behavior of individual components which make up the interferometer.  Phases delays, polarization rotation, and attenuation will be identical for reciprocal waves passing through a reciprocal component.

The second simplifying principle is that with a broad-band source, interferometric fringes are visible only between modes with optical path length difference less than the source's coherence length\cite{lefevre}.  Because of the strong birefringence of the PM fiber and EOM, most scattered modes will have path lengths which differ by many times the coherence length.  Light following these ``incorrect" paths will reach the detector, but won't interfere coherently with the two main modes, and therefore can not generate a spurious signal.  These modes will, however, shift power away from the interfering modes, and will appear as increased signal at the ``DC" output of the detector.  The ratio of the power on the DC part of the signal to the power on the second harmonics is a good measure of scattering as indicated by deviations from the theoretical prediction of Equation \ref{expansion}.  By this indicator, we note that the all-fiber interferometer possesses more scatter than the free-space SI.

\begin{figure} [h]
\includegraphics[width=.80 \columnwidth]{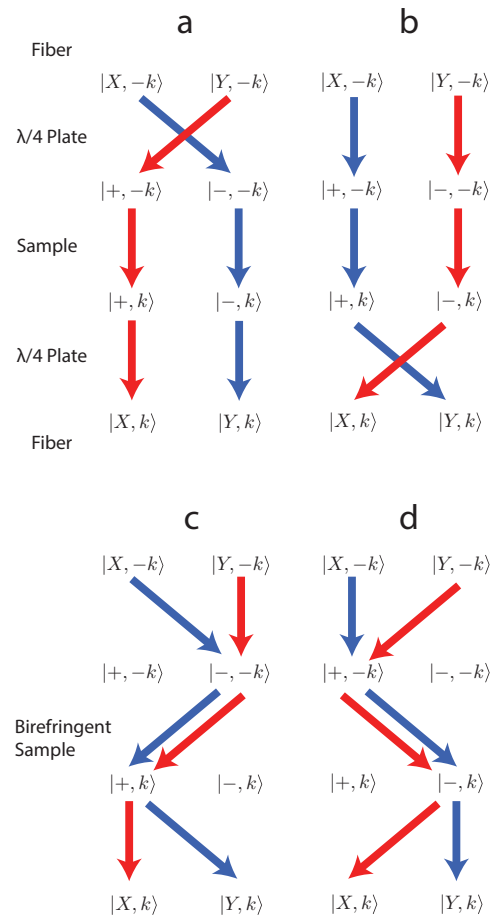}
\caption{Illustration of all eight possible mutually coherent scattering modes, grouped in reciprocal pairs (red/blue), leaving the fiber and interacting with the sample.  All of these could affect to the measurement of the Kerr signal, since the sum of the four amplitudes per channel, yields the total transition amplitude which is then interfered.  The intermediate polarization and momentum basis states are written in the ket notation.  (a) and (b) are those modes that appear alone for the two ideal angular positions of the the quarter-wave plate.  c and d are those modes added when there is birefringence in the sample.  The $\pm$ defines the circular polarization in terms of the spin angular momentum state of the field}  
\label{reflectionModes}
\end{figure}

The analysis of the interferometer is considerably simplified by the spatial filtering provided by the polarization maintaining fiber as it allows for a description of the electromagnetic state by only two polarization states at all points in the instrument.  It does not matter that the beam incident at the sample is not a plane wave; there are still two basis states of the field, both of which possess an approximate Gaussian beam profile, have opposite spin angular momentum and which are respectively mode-matched to the two linear polarization states within the fiber.  Even if the quarter-wave plate is imperfect and the outgoing states are only elliptically polarized, because the incoming polarization states are both linear and orthogonal, the expected value of spin angular momenta (not including power differences between the beams) of each mode leaving the quarter-wave plate will be equal and opposite.

In Figure \ref{reflectionModes}, we consider intermediate basis states of the two modes of the optical field as they pass through the various optical components and the reflecting sample.  At each transition, the basis has been defined so that in a perfectly aligned system, only a single basis element describes the optical state of each mode at every point in the fiber.  In an imperfectly aligned system, superpositions of these modes must be considered.  Ideally, the modes leave the fiber linearly polarized, $\vert X/Y,-k\rangle$, pass through the quarter-wave plate and become circularly polarized as $\vert \pm,-k\rangle$.  After reflecting off the sample ($\vert \pm,k\rangle$), the quarter-wave plate returns the states to the linear polarization, $\vert X/Y,k\rangle$, and couple into the fiber.  Since only light that traverses from X to Y polarization and vice versa will be coherent and interfere, only such modes have to be considered in the analysis.

In the figure, the eight relevant paths are grouped in reciprocal pairs, so by the reciprocity theorem, the contribution to the full transition amplitudes from each of these pairs of paths will be identical if all the optical components are reciprocal.  When the the quarter-wave plate is aligned perfectly, the modes illustrated in either only Figure \ref{reflectionModes}a or only in \ref{reflectionModes}b will be occupied.  Thus, there are two possible angular positions of the quarter-wave plate, $90^\circ$ apart, for which the interference fringe visibility is optimized.  Switching between these two orientations exchanges the phase-delays accumulated upon reflection between the two channels.  As it is the phase difference between the two modes that is measured, the Kerr angle signal will change sign.  When there are misalignments and imperfections in the quarter-wave plate so that the incident light is elliptically polarized, then those paths represented in both a and b will be present and coherent at the detector.

In general, we have $\langle X,k\vert Y,-k\rangle_b\propto\langle Y,k\vert X,-k\rangle_a\propto e^{-i\theta_K}$ and 
$\langle X,k\vert Y,-k\rangle_{a}\propto\langle Y,+k\vert X,-k\rangle_b\propto e^{i\theta_K}$.  Here, the constant of proportionality is real and we are abusing the bra-ket notation, with subscript (a,b,c,d) to denote the product of transition amplitudes between each basis state constituting the respective path in Figure \ref{reflectionModes}.  So, for each element of the transition matrix for propagation outside the fiber, we have, for example, $\langle X,k\vert Y,-k\rangle=\sum_u \langle X,k\vert Y,-k\rangle_u$ and $\langle X,k\vert Y,-k\rangle_a=\langle X,k\vert +,k\rangle\times\langle +,k\vert +,-k\rangle\times\langle +,-k\vert Y,-k\rangle$.  When there is birefringence in the sample, then those paths illustrated in Figure \ref{reflectionModes}c and \ref{reflectionModes}d become relevant.  From the figure, it is apparent that the sample's contribution to each pair of amplitudes is the same.  Likewise, will be the contribution from the quarter-wave plate and the fiber, as they are reciprocal components.   Consequently, the full amplitudes for each pair will be identical regardless of whether the sample is reciprocal or not: $\langle X,k\vert Y,-k\rangle_u=\langle Y,k\vert X,-k\rangle_u$, for $u=c,d$.  

When some or all of these paths are present, they will interfere coherently at the detector because the optical path length will be the same for all of them.  Thus, if the sample is non-reciprocal, then the instrument will return an incorrect reading for the Kerr angle.  If the sample is reciprocal, then all reciprocal paths will contribute the same amplitude to both channels and no Kerr signal will be measured.  Consequently, misalignments, imperfections, birefringence or any other reciprocal perturbation to the system will not introduce spurious signals.

When the QWP is removed entirely, light will mostly follow \ref{Paths}c, but any amount of scattering, especially at the fiber facet, will give way to optical modes following the ideal paths of \ref{Paths}a, and be sensitive to magneto-optical effects.  We measure the magnitude of the second harmonics when the  quarter-wave plate is removed to be $5\times 10^{-3}$ smaller than when it is present and correctly aligned.  Because the scattering that gives rise to rays following path \ref{Paths}a with the quarter-wave plate removed is functioning as a quarter-wave plate, upon adding a real quarter-wave plate back into the system, these scattering modes will follow the paths indicated by \ref{Paths}c, and so will be incoherent with the primary paths and not contribute to the Kerr signal.

There are two points in the system, as illustrated in Figure \ref{Paths}d, where back-scatter into the same channel will result in beams of light which have the same path length as the primary interfering modes, and will therefore interfere.  To minimize the possibility of scattering at these points, the second customization requirement for the inline EOM maximizes the path length difference between the fast and slow modes upon exiting the fiber at the sample.  This guarantees that the two back-scatter positions are as far away from terminating facet of the PM fiber as possible, thus avoiding cross-coupling from scattering at the fiber facet surface.  As the path length difference between the modes leaving the fiber is $\Delta\ell=L\times(n_e-n_o)=9.2 mm$, one of the back-scatter points is located about $\tfrac{\Delta\ell}{2n_e}=3 mm$ behind the fiber terminus and the other point is outside the fiber within the focusing lens, where back-scattering is unlikely to re-couple into the fiber due to mode-mismatch.  Since any scattering surfaces or any other sources of cross coupling are located a distance greater than the 8 $\mu$m coherence length of the light from any of these back-scatter positions, only Rayleigh scattering from impurities can possibly generate the unwanted modes.

\section{Initial Tests of the Apparatus}

\begin{figure} [h]
\includegraphics[width=1.0 \columnwidth]{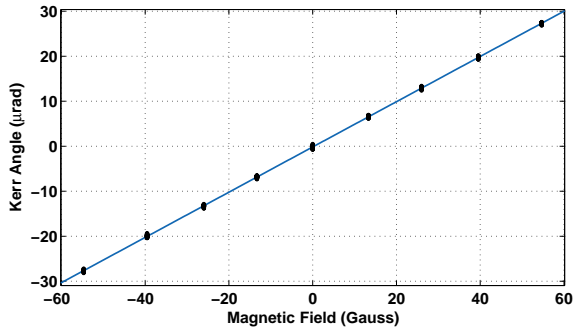}
\caption{Faraday rotation measurement through a .5 mm thick sample of quartz with one side coated with Aluminum.  As the beam passes through the sample twice, the effective width of the sample is 1 mm.}
\label{quartz}
\end{figure}

To demonstrate the accuracy of the instrument, we measure the Verdet constant of a 0.5 mm thick sample of z-cut quartz.  One side is coated with a reflective layer of aluminum, so the light makes two passes through the quartz before returning to the fiber.  Figure \ref{quartz} is the Faraday rotation from the double pass measured as a function of magnetic field which gives 0.50 $\tfrac{\mu\text{rad}}{G}$.  For reference, we flipped the sample over and measure the Kerr rotation reflecting off just the Aluminum layer: 0.27 $\tfrac{\mu\text{rad}}{G}$.  The Verdet constant is given by the difference 0.23 $\tfrac{\mu\text{rad}}{G mm}$,  which is compares well with a previous measurement by Ramaseshan\cite{ramaseshan} of 0.26 $\tfrac{\mu\text{rad}}{G mm}$.  In both measurements there was a $-0.15 \mu$rad offset at zero field, which is likely due to RAM.

\begin{figure} [h]
\includegraphics[width=1.0 \columnwidth]{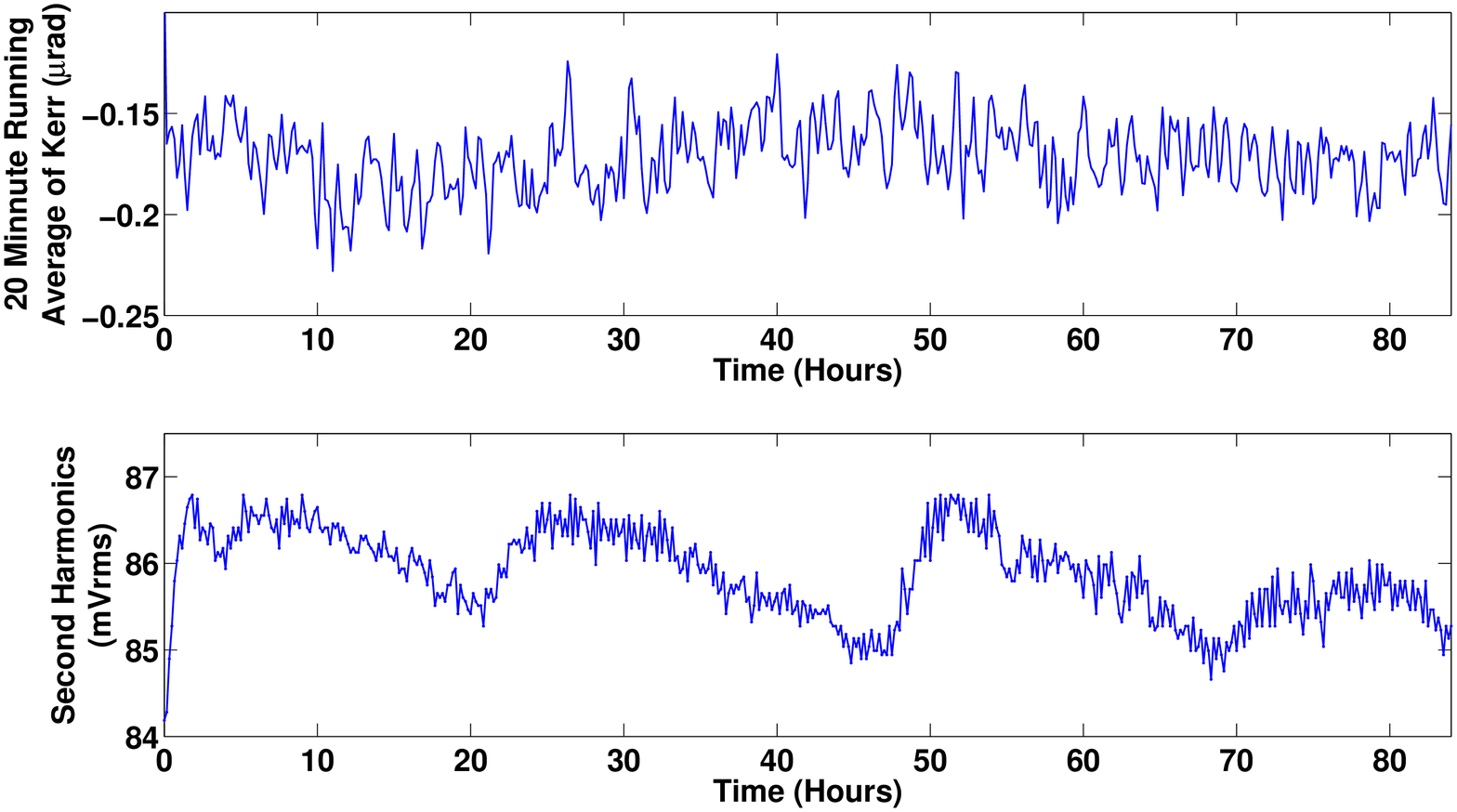}
\caption{Drift of the Kerr angle (a) and the Second Harmonics (b) of a gold mirror over time.  Although there are drifts in the Second Harmonic signal, the Kerr signal is unaffected.}
\label{drift}
\end{figure}

\begin{figure} [h]
\includegraphics[width=1.0 \columnwidth]{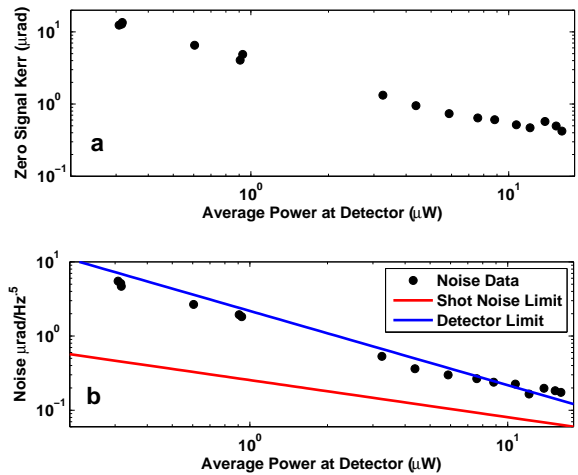}
\caption{(a) Spurious offset and (b) the rms noise of the Kerr signal off a gold mirror as the power returning to the detector is varied by defocussing the sample from the focal point.  The estimated noise figures are given by Equations \ref{shotN} and \ref{detecN}.  Do to the large power loss in the EOM the all-fiber interferometer is detector-noise limited.}
\label{noisefig}
\end{figure}

Much of SI's performance can be characterized by studying how the measurement varies as the power varies.  We tune the power received by the detector by moving the lens out of focus from the sample, rather than changing the power through the SLED, as this can affect its spectrum, and therefore change how the performance of the SI, which is nominally aligned only for a single optical frequency.  Noise, as measured by the standard deviation in the Kerr signal is well described by thermal detector noise, as can be seen in Figure \ref{noisefig}b.  At higher powers, the shot noise will dominate; the two noise sources are distinguished by their power law behavior and this is discussed in detail in Appendix \ref{calculations}.    

Another of the standard measures of performance for SIs is how much the signals drift over long periods of time.  These drifts can enter from thermal fluctuations within the EOM, air currents and re-amplified back-reflections to the SLED.  In Figure \ref{drift}, the Kerr signal does not drift appreciatively over about 84 hours, although in practice the drift does not change over the course of many months.  That the Second Harmonic signal does exhibit drift suggests that these variations is successfully canceled out in the division operation when the Kerr signal is calculated.

Finite offsets to the Kerr signal are another possible way in which the SI may give false readings, and although stray signals of up to a $\mu$rad have been recorded, the present alignment, as shown in Figure \ref{drift}a exhibits an offset of $\sim .18 \mu$rad, measured over Gold.  Gold is not expected to yield a Kerr effect in the absense of an applied field and the fact that the observed offset changes between various alignments of the system suggests that the source is instrumental.

We have found that these anomalous, yet stable, offsets originate from stray signals coupling to the electronics at the EOM drive frequency and from RAM.  When it is the former, the offset signal on the first harmonics will not depend on power; this results in a false finite measurement of the Kerr signal: $(\phi_{nr}\times \mathcal{I}+\delta)/\mathcal{I}$.  As the power decreases, the presence of the $\delta$ offset will generate a larger false value for the Kerr angle in the calculation as shown in Figure \ref{noisefig}.  RAM can be discerned from a Kerr signal, as the sign of the measured value will not change when the quarter-wave plate is rotated $90^\circ$.  We discuss this in detail within Appendices \ref{RAMap} and \ref{alignment}.  

\section{The Scanner}

\begin{figure} [h]
\includegraphics[width=1.0 \columnwidth]{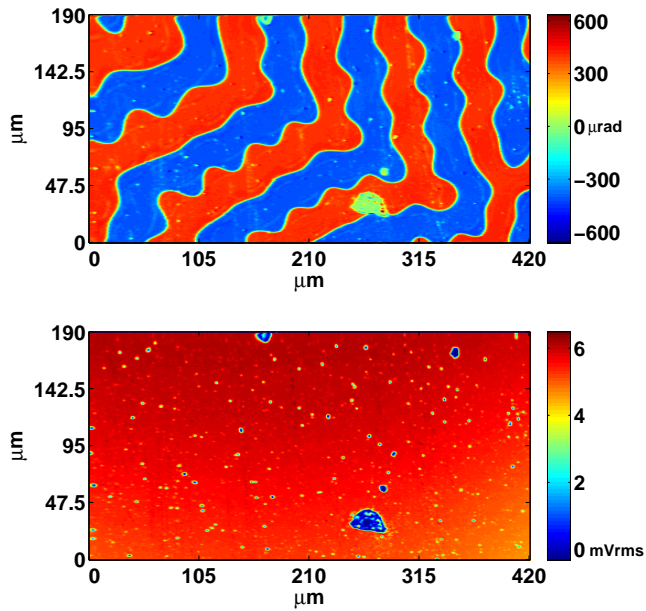}
\caption{Scan of Bismuth Iron Garnet at room temperature.  (a) is a rightward moving trace of the Kerr signal and (b) is the corresponding Second Harmonic amplitude, which is a proxy for reflectivity, though birefringence should also decrease its value\cite{hovo}.  The values of the Kerr angle on the domains are $\pm$501 $\mu$rad.}
\label{garnet}
\end{figure}

A homebuilt 2D scanner and driving circuit was constructed to affect translation of the optical fiber at cryogenic temperatures.\cite{wynnthesis,siegel}  Two pairs of piezo bimorph ``S-benders" allow translation with minimal change in the sample-to-scanner distance over approximately a 1 mm$^2$ area at room temperature.  An Attocube nanopositioner allows for vertical motion of the sample to allow for easy focusing of the beam.  The scanner's voltage response was calibrated with a gold bar of width 8.2$\mu$m.  To demonstrate the imaging capabilities of the all-fiber interferometer scans of a 320 um thick Bismuth Iron Garnet crystal from Integrated Photonics  was examined at room temperature as shown in Figures \ref{garnet}a and \ref{garnet}b.  The domain widths are about 20 $\mu$m wide and have uniform magnetization with a sharp transition at the domain wall.  The second harmonics suggest that the sample was slanted slightly with respect to the beam and so moved slightly out of focus, decreasing the amount of light re-coupled to the fiber at points in the lower right-hand corner of Figure 6b.  Dirt on the surface is clearly visible and it possesses no magnetization.

\begin{figure} [h]
\includegraphics[width=1.0 \columnwidth]{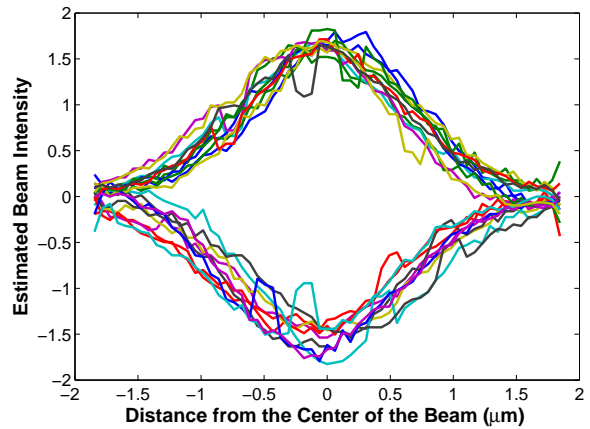}
\caption{The beam profile is estimated by taking the discrete derivative of the second harmonic trace with respect to the displacement of the scanner as the beam is scanned back and forth across an Au/GaAs interface many times (a subsequent Abel Transform is an over-refinement for profiles that are approximately Gaussian).  As the sign depends on the direction of motion, positive values of the derivative are from rightward-moving traces, while negative values are from leftward traces.  There is hysteresis in the scanner so each trace above has been displaced so the maximum/minimum is at 0 $\mu$m.}
\label{beam}
\end{figure}

In Figure \ref{beam} we show the profile of the beam as measured by scanning across the edge of a gold bar and taking the difference of the second harmonic magnitude at adjacent points.  This shows that the spot size is about 1.5 $\mu$m in diameter, which agrees with diffraction limit. The lens has a focal length of about 1.1 mm and the diffraction limited spot size allows the system to be robust against angular misalignment of the sample.  However, this makes it sensitive to mechanical shocks which induce vibrations in the z-direction.  We find that these vibrations damp out quickly and primarily register on just the second harmonic signal, a tribute to robustness of the measurement against intensity fluctuations.

\section{Conclusion}

The SI is an elegant method to measure weak magnetization.  To further emphasize the flexible nature of this technique we conclude by mentioning several alternative imaging modes.  Replacing the quarter-wave plate with a $\pi$/4 Faraday rotator will rotate the orthogonal, linearly polarized modes from the fiber by $45^{\circ}$.   Upon reflection from the sample, those modes will, again, be rotated by $45^{\circ}$  by the Faraday rotator--for a net rotation of $90^{\circ}$.  In this way, polarization states within the fiber will again interchanged and the optical paths will form a similar reciprocal zero-area loop to the above.  The only non-reciprocal phase shifts in this configuration will come from linear birefringence in the sample and would, essentially, measure anisotropy.  Another possible application is that natural gyrotropy in conductive materials may be sensed by taking advantage of the inverse circular photogalvanic effect, in which a current propagating along a screw axis may generate a weak spin-polarized current that would appear as a magnetic signal with sign dependent upon the direction of the current.\cite{ivchenko}  Finally, the system may be operated as a magnetometer by simply imaging a reflective material with high Verdet-constant in close proximity to the sample\cite{xiathesis}.

\section{Acknowledgments}
Very special thanks to Hovnatan Karapetyan, Elizabeth Shemm, Jing Xia, Min Liu, Alan Fang, Zhanybek Alpichshev, Carsten Langrock, Jason Pelc, Clifford Hicks and Beena Kalisky.  We also thank Alexander Palevski, Axel Hoffmann, Helmut Schulthei\ss, Robert Feigelson, and Mark Randles for providing samples for calibration.  We are grateful for support from the Center for Probing the Nanoscale, Grant NSF NSEC Grant PHY-0830228.

\appendix
\section{Extracting the Kerr Signal}
\label{calculations}
As the function of most of the components in the system is to ensure propagation of the two modes containing the magnetic phase information, without cross-coupling, the behavior of the apparatus is modeled by considering contributions to the phase delay from the EOM and the magnetic sample only.  Without loss of generality, we can assume that the EOM's phase-modulation action occurs only along the TM mode.  Below, we expand upon similar discussions in references.\cite{xiathesis,dodgethesis}  At the detector, the optical signal is: 

\begin{equation}
I(t)=\tfrac{1}{2}\mathcal{I} \lvert e^{i\phi_m\text{sin}(\omega t) +\theta_K} \pm e^{i\phi_m\text{sin}[\omega (t+\tau_{path})] -\theta_K} \rvert^2
\label{intensity}
\end{equation}

The two terms within the norm above correspond to the phases accumulated on the two counter-propagating modes in Figure \ref{sagnac}.  $\mathcal{I}$ is the power.  $\phi_{m}$ is the phase modulation depth of EOM as set by the driving voltage.  $\omega$ is the driving frequency as set by the function generator such that $\omega \tau_{path}$  is the phase delay introduced by the transmission through a double pass through the $\sim$10 m of fiber, and the $\theta_K$ phase shifts are those generated by the nonreciprocity within the sample.  In general, these phases are not equal and opposite, as the reciprocal components will add a fixed phase to both beams, but since the instrument measures only the difference in phases, this is irrelevant.  As illustrated in Figure \ref{reflectionModes}, there are two angular positions of the quarter-wave-plate which align the system, and these determine if a factor of $-1$ is included in the measurement of the Kerr signal.  We have calibrated our instrument so that the terms are added.  Thus, light leaving the slow axis from the end of the fiber is $+$ circularly polarized as in Figure \ref{sagnac} and that positive Kerr signal implies that magnetization is out of the sample's surface.

The optimal performance of the interferometer occurs when the fiber length and frequency are selected to provide a $\pi$ phase shift on the second pass through the modulator: $\omega c[(n_e+n_o) L +\tfrac{1}{2}(N^{EOM}_e+N^{EOM}_o)L_{EOM}] =\pi$, where $L\approx 10 m$ is the fiber length and $n_{e/o}\approx 1.458$ are the indices of refraction for the extraordinary and ordinary axes within the fiber (birefringence is $3.5\times 10^{-4}$), and $N^{EOM}_{e/o}$ and $L_{EOM}$ are the indices of refraction of the two axes in the EOM and it's length.
By the Jacobi-Anger expansion:
\begin{align*}
e^{i z \cos \theta}&=\sum_{n=-\infty}^{\infty} i^n\, J_n(z)\, e^{i n \theta} \\
e^{iz \sin \theta} &= \sum_{n=-\infty}^\infty J_n(z) e^{in\theta}
\end{align*}

Where $J_n(z)$ is the n$^{\text{th}}$ Bessel Function.  Equation \ref{intensity} thus yields:
\begin{equation}
\begin{split}
I(t)&=\mathcal{I}\left(\tfrac{1}{2}\left ( 1+J_0(2\phi_m)\right )+ \text{sin}(2\theta_K)J_1(2\phi_m)\text{sin}(\omega t) \right). \\ 
 & \left.+ \text{cos}(2\theta_K)J_2(2\phi_m)\text{cos}(2\omega t) + ... \right )
\label{expansion}
\end{split}
\end{equation}

The Kerr angle is calculated from the harmonic coefficients in Equation \ref{expansion}: 

\begin{equation}
\theta_K=\frac{1}{2}\text{tan}^{-1}\left[\frac{J_2(2\phi_m)V_{1\omega}}{J_1(2\phi_m)V_{2\omega}}\right ]
\label{kerr}
\end{equation}

Where $V_1$ and $V_2$ are the rms voltage outputs from the Lock-Ins measuring the first and second harmonics respectively.  Since these voltages above are divided against each other, variations in the power, either from a noisy fluctuating SLED or changes in the sample's reflectivity will not contribute to the measurement as all harmonics will be affected equally.  Strong decreases in reflectivity will affect the signal noise, however.

The noise is estimated from Equation \ref{kerr} by relating the noise voltages, $\delta V$, to the noise power $\delta \mathcal{I}$ for each noise source:
\begin{equation}
\theta_K=\frac{1}{2}\frac{J_2(2\phi_m) \delta V_{1\omega}}{J_1(2\phi_m)V_{2\omega}}=\frac{1}{2}\frac{J_2(2\phi_m) \delta \mathcal{I}_{1\omega}}{J_1(2\phi_m)\mathcal{I}_{2\omega}}
\label{noise}
\end{equation}

The noise for this measurement comes from two primary sources:

\begin{equation}
\begin{split}
&\text{Shot noise:} \\
&\Delta \theta_K=\frac{1}{2}\frac{J_2(2\phi_m)}{J_1(2\phi_m)}\frac{\sqrt{2(hc/\lambda)\times \eta \times EBW \times P_{dc}}}{s\times P_{dc}}
\end{split}
\label{shotN}
\end{equation}

and
		 
\begin{equation}
\begin{split}
\text{Detector noise:}& \\
&\Delta \theta_K=\frac{1}{2}\frac{J_2(2\phi_m)}{J_1(2\phi_m)}\frac{NEP}{s\times P_{dc}}
\end{split}
\label{detecN}
\end{equation}

$\eta$ is the quantum efficiency of the detector, NEP is the noise equivalent power of the detector, and EBW is the Effective noise Bandwidth of the lock in amplifier as set by its internal filter and the time constant.  The constant, $s\sim1$, is a measured proportionality factor that relates the voltage on the DC part of the signal to RMS voltage on the second harmonics.  

The Sagnac interferometer is a direct probe of reciprocity because it interferes light propagating along the two reciprocal channels of in Figure \ref{sagnac}.  Direct application of the reciprocity theorem of classical electromagnetism to this system guarantees that only nonreciprocal effects will cause the amplitude and phase of the two optical paths to be different after traversing the loop.  Aside from the sample, and the EOM, which introduces a time dependent bias in the controlled manner shown above, all of the other optical components along the path are non-magnetic, and thus, reciprocal.  Provided that the procedure below is followed, even in the presence of misalignments or imperfect components, the SI will measure only non-reciprocal effects in the sample. The robustness of Equation ~\ref{intensity} to imperfect components and other variations in the model can be demonstrated explicitly by employing the Jones calculus, however the principle of reciprocity provides a simple and accurate intuition.

Strong magnetic fields applied to the fiber or the lenses and along the optical propagation direction will also cause nonreciprocal phase shifts.  In the case of a Faraday effect within the fiber, the two interfering paths will travel through the applied field in both directions.  The phase shift on a linear polarization state from a magnetic field can not depend upon the direction of propagation, so both beams will yield the same phase and there will be no net phase difference.  For this reason, it is important to place the quarter-wave plate between the lens and the sample.  Otherwise, if there is Faraday rotation in the lens from an applied magnetic field, opposite circularly polarization states will yield different phase delays and this will contribute to the Kerr signal.  When the quarter-wave plate is placed after the lens, then the  linearly polarized states will still exhibit Faraday rotation, but because the rotation will be identical for both beams after traversing the lens twice, there will be no net phase difference.  Nonlinear magneto-optical effects will not exhibit this cancellation and we observe reciprocity for magnetic fields orthogonal to the propagation direction.  This could be a consequence of the Cotton–Mouton effect amongst other magneto-optical effects\cite{landau}.

A more explicit understanding of the workings of the SI can be gained by application of the Jones Matrix formalism.  However, this must be used  carefully in the case of a broadband source, as it is necessary, especially when modeling misalignments and imperfections, to perform the sum in Equation \ref{intensity2} over all wavelengths, so as to take into account amplitudes which are incoherent and will not interfere.  In the ideal alignment, calculated  below, the sum over optical frequencies, k, is superfluous, because there is no weight on amplitudes corresponding to the improper paths.  For convenience, we supply the relevant Jones matrices for the optical components illustrated in Figure \ref{sagnac} and they can be modified without discretion to explicitly test for non-reciprocity when modeling the effects of imperfections.

\begin{center}
   \begin{tabular}{p{1.5in}l} 
Polarizer: & $ P=\begin{bmatrix}1 & 0\\
0 & 0
\end{bmatrix} $\\
Intensity: &$\rho_k=\begin{bmatrix}\mathcal{I}_{e}(k)\\
 & \mathcal{I}_{o}(k)
\end{bmatrix}$\\
 Beam Splitter at Port 1 of the EOM:  &$B=\frac{1}{\sqrt{2}}\begin{bmatrix}1 & 1\\
1 & -1
\end{bmatrix}$\\
EOM: &$E(t)=\begin{bmatrix}e^{i\phi_{m}(t)} & 0\\
0 & 1
\end{bmatrix}$\\
$\frac{\lambda}{4\,}$ Plate:  &$Q=\frac{1}{\sqrt{2}}\begin{bmatrix}1 & i\\
i & 1
\end{bmatrix}$\\
$\frac{\lambda}{4\,}$ Plate (rotated $90^\circ$):  &$Q'=\frac{1}{\sqrt{2}}\begin{bmatrix}1 & -i\\
-i & 1
\end{bmatrix}$\\
Fiber: & $F_k=\begin{bmatrix}e^{in_{e}kL} & 0\\
0 & e^{in_{o}kL}
\end{bmatrix}$\\
Sample: & $M=\begin{bmatrix}\cos\theta_{K} & \sin\theta_{K}\\
-\sin\theta_{K} & \cos\theta_{K}
\end{bmatrix}$\\
   \end{tabular}
 \end{center}

The trace operation is a succinct method for calculating the modulated intensity observed at the detector:

 \begin{equation}
\begin{split}
I(t)=&\Bigg|\underset{k}{\sum}\text{Tr}\Big[\rho_k \times P\times B\times E(t+\tau)\times F_k\times Q\\
&\times M\times Q\times F_k E(t)\times B\times P\Big]\Bigg|^{2}
\label{intensity2}
\end{split}
 \end{equation}

which yields Equation \ref{intensity}.

\section{Models and Measurements of Residual Amplitude Modulation}
\label{RAMap}

The Sagnac Interferometer exhibits a finite offset to the Kerr signal.  One possible origin of this offset is from Residual Amplitude Modulation from the EOM and its contribution to a spurious Kerr signal can always be distinguished from those originating from the non-reciprocal signals as its sign will not change when the quarter-wave plate is rotated $90^{\circ}$.  RAM can add spurious harmonics to the signal in many ways, but because of the reciprocal nature of the interferometer, many of these are canceled out.  Though we are unable to fully account for the size of the offset, we can model some of the more qualitative features and provide understanding for methods of reducing spurious signals.  

In addition to the fiber-position dependence, mentioned earlier, we provide some additional information regarding its behaviors.  The offset does not vary when different samples are imaged.  It also is not from an RF electronic noise signal inadvertently coupling to the detector or the lock-in amplifiers as the spurious Kerr signal does not depend on intensity, nor is it present when the beam is blocked at the sample.  It is also not likely to be from stray magnetic fields, as none have been detected near the instrument with a gauss-meter.  It is not from a spurious intensity modulation present on the SLED, as might be induced from stray coupling from the function generator, as no signal is detected when the driving field on the EOM is removed.  Finally, there is a wavelength dependence to the offset, which suggests that the optical path length is an important part of the effect.  The effect is also different when the system is dismantled and then re-assembled.  Specifically, if the fiber-connectors between the EOM and the sample are slightly misaligned, the spurious amplitude modulation increases.  The RAM also depends upon the layout of the fiber: when it is coiled on its spool there may be no offset, but when the whole fiber is stretched out or twisted, stable offsets on order of a 1 $\mu$rad occasionally appear.  The only way to eliminate this is by moving the fiber into a position where the offset is negligible.  Some of these characteristics suggest that the non-ideal modes of Figure \ref{Paths} might be contributing to the spurious signals, and that scattering and variations in the index of refraction within the fiber are the source.

We measure the size of the RAM at the lock-in frequency for light after a single pass through the inline fiber EOM to be about $\alpha=10^{-5}$ times the average power.  The exact value is stable, but it changes every time the SLED power is adjusted.  The variations are greater than by no more than a factor of about 2.  We also measure the phase of this signal with lock-in relative to some constant phase, though the exact value can be estimated from the cable length to the detector.  While there appears to be a hysteretic effect on the RAM when SLED power is varied, the value of the corresponding phase measurement is independent of power.  We also measure RAM at the second harmonic frequency, but and this is decreased in magnitude by about a factor of 4 from that on the first harmonic.

\subsection{Harmonic Expansion}
There are many optical trajectories in the system if scattering and improper couplings are considered as shown in Figures \ref{Paths} and \ref{reflectionModes} and any of these might contribute to spurious signals.  As mentioned before, we find that only those paths which include reflection off the sample can possibly generate an offset, as no signal is observed when the sample is removed.  The most general model for RAM included on the two primary paths is:

\begin{equation}
\begin{split}\label{RAM0}
I(t)=&\tfrac{1}{2}\mathcal{I}\lvert A_1(t)e^{i\Phi(t)+i\theta_K} + A_2(t)e^{i\Phi(t+\tau_{path}) -i\theta_K} \rvert^2\\
\Phi(t)&=\phi_m\text{sin}(\omega t)\\
A_1(t)&= (1+\underleftarrow{\alpha_o} \text{sin}(\omega (t+\tau_{path})))\times (1+\underrightarrow{\alpha_e}\text{sin}(\omega t)))\\
A_2(t)&= (1+\underleftarrow{\alpha_e} \text{sin}(\omega (t+\tau_{path})))\times (1+\underrightarrow{\alpha_o}\text{sin}(\omega t)))
\end{split}
\end{equation}

$\alpha_e$ and $\alpha_o$ distinguish the modulation applied to the light on the extraordinary and ordinary axes of the EOM when it is moving in the direction indicated by the underscored arrow.

Consider the situation where there is no directional dependence to the EOM's amplitude modulation.  If $\tau_{path}$ is optimized to yield a $\pi$ phase shift for a complete traversal of light around the Sagnac loop, then the amplitude of odd harmonics of the modulation will flip sign after a second pass through the EOM.  Using this condition, we study the most general model of RAM where it has spectral weight on all integral harmonics of the drive frequency.  Let $F_n(t)$ and $S_n(t)$ be equal to the complex harmonic (i.e. $F_n(t)=F_{-n}^*(t)=a_n\exp(i n\omega t+i\theta_n)$) of RAM with arbitrary phase and amplitude at frequency $n\omega$ on the fast and slow modes of the EOM respectively. 
\begin{align*}
A_1(t)&= \overset{\infty}{\underset{n=-\infty}{\sum}} F_n\times \overset{\infty}{\underset{n'=-\infty}{\sum}} (-1)^{n'} S_{n'}\\
A_2(t)&= \overset{\infty}{\underset{n=-\infty}{\sum}} (-1)^{n} F_n\times \overset{\infty}{\underset{n'=-\infty}{\sum}}  S_{n'}\\
\end{align*}
Where the sums above are taken over the harmonics of the driving frequency.  Expanding \ref{intensity} with this included
\begin{equation*}
I(t)=\tfrac{1}{2}\mathcal{I}\left( A_1^2(t)+A_2^2(t)+A_1(t)A_2(t)\times\text{J-A Expansion}\right)
\end{equation*}
So,

\begin{align}
\begin{split}\label{Ram1}
&A_1^2(t)+A_2^2(t)=\\
&\sum_{nn'n''n'''} F_\omega F_{n'} S_{n''} S_{n'''} \left((-1)^{n+n'}+(-1)^{n''+n'''}\right)\\
\end{split}\\\label{Ram2}
&A_1(t)\times A_2(t)=\sum_{nn' n''n'''} F_n F_{n'} S_{n''} S_{n'''} (-1)^{n+n''}
\end{align}

A term will contribute to the $n^{th}$ harmonic of these terms if $\vert n + n' + n'' + n'''\vert=b$.  If $b$ is odd, then an odd number of mixing frequencies will be odd harmonics.  Of the four frequencies above, dividing them into any two pairs will require one set to sum to an odd number, and the other set to an even number.  This implies that there are no odd harmonics in both of the above terms.  When there is no Kerr signal, then there are only even harmonics in the Jacobi-Anger expanded interference terms, so any mixing between them and the RAM will produce only even terms.  To conclude, there must be a drive frequency at which amplitude modulation introduced in this fashion can not contribute to the first harmonics in an aligned system.  Spurious even harmonics must be at most at the level of $\alpha$.  

If the frequency is misaligned then the spurious signals will be on order $\delta\times\alpha$, where $\delta=\pi\frac{f-f_{ideal}}{f_{ideal}}$.  A slight misalignment in the frequency approximately involves replacing (-1) with $\exp(i\pi-i\delta)\approx(-1+i\delta)$ in Equations \ref{Ram1} and \ref{Ram2}, and expanding to first order in $\delta$.  Spurious harmonics with magnitude comparable to a 1$\mu$rad signal will occur when $\delta\times\alpha\approx 10^{-6}$ or when $\delta\approx 10^{-1}$, and since the precision of the frequency set-point is on the scale of $10^{-3}$, misalignments in frequency are unlikely to generate a noticeable offset.

Despite precautions, such as angle polished fibers, additional, unintended modes can be present in the interferometer having formed by scattering or reflections, such as in Figure \ref{Paths}d.  If two such modes have the same optical path length, they will form a competing interferometer with signal superimposed on top of the primary one \ref{Paths}a at the detector.  With exception given to those modes shown in Figure \ref{Paths}d, It is highly unlikely that such modes comprising a spurious interferometer have a path length identical to the primary modes.  Consequently, they are driven by the EOM at a frequency not equal to $\tfrac{2c}{\text{Path Length}}$ and as mentioned above, this will generate an offset from RAM.  We have noticed an increase in RAM when fiber connectors are not well aligned.  Such misalignments, as well as others, might introduce modes which form additional interferometers.

\subsection{Directional Asymmetry in Traveling-Wave Modulators}

The analysis above demonstrates that the effects of directionally independent RAM should be tuned away with proper drive frequency alignment for the EOM, however we still observe a finite offset on the order of a $\mu$rad to the Kerr signal.  We consider the case that there is directional dependence to the modulation coefficients in Equation \ref{RAM0} and demonstrate that it, too, can not account for the observed spurious amplitude modulation.  Directional dependence might be expected given that the waveguide modulator operates by applying a propagating RF electric field to the Lithium Niobate crystal that travels at the same speed as the TM (extraordinary) optical polarization component moving in the direction away from the circulator in Figure \ref{sagnac}.  The relative difference in the propagation speeds of the two waves is less than 10\% from factory specifications, however when the optical beam is traveling in the direction opposite to that of the RF field, the velocity mismatch will be greater.  Because of the directional dependence of the phase modulation, it is to be expected that there will be a directional dependence to the amplitude modulation as well.  $F_\omega$ and $S_\omega$ will be directionally dependent and the spurious first harmonics will not cancel in Equations \ref{Ram1} and \ref{Ram2}.  

We consider a model of the traveling wave modulator where the action of the modulator on the optical beams not only depends on polarization, but on the propagation direction as well.  As a function of position and time, the voltage at any point in the traveling wave modulator is given by\cite{wavemodulator}:
\begin{equation*}
V(z,t)=V_0 \text{sin}(k N_m z -\omega t)
\end{equation*}
where $N_m\equiv\tfrac{c}{v_{\text{RF}}}$ is the propagation constant within the electrodes for the driving voltage signal, $k$ is the free space RF wave number and $\omega$ is the frequency of the RF signal.  This equation assumes that the wave starts on one side of the Lithium Niobate crystal waveguide, of length L, and is absorbed at the other end.  As the modulator used in these experiments is not terminated with a 50$\Omega$ resistor that would otherwise allow for impedance matching, part of the signal will be reflected and a corresponding term for this traveling wave will need to be included in the above equation.  We will consider perfect impedance matching and extend the argument to the generalized situation.  If the optical signal is propagating with the RF wave then its position, $z$, is given by $k N^{EOM}_{e/o} z=\omega (t-t_0)$, where $N^{EOM}_{e/o}$ is the index of refraction for the extraordinary or ordinary optical polarization states, and $t_0$ is the time when the light beam first enters the EOM from either direction.  The phase shift, as a function of position, will be proportional to the applied voltage by $\beta_{e/o}$, which depends on the electro-optic properties of the crystal.  A propagating optical beam will have a phase shift per unit length within the modulator given by:
\begin{align}
\begin{split}
\Delta\phi_{e/o}& (z)=\beta_{e/o} V_0 \sin(k (N_m-N^{EOM}_{e/o}) z -\omega t_0)\\
\intertext{and it's integral over the entire modulator is:}
\phi^\rightarrow_{e/o}&= \int_0 ^L \Delta\phi_{e/o}(z) dz\\
&=\beta_{e/o} 2\frac{V_0}{k \Gamma_{e/o}} \text{sin}\Big(\frac{k \Gamma_{e/o} L}{2}\Big)\text{sin}\Big(\frac{k\Gamma_{e/o} L}{2}-\omega t_0  \Big)\\
&= \beta_{e/o} V_0 L (1+\underset{\rightarrow}{\epsilon})\text{sin}\Big(\frac{k\Gamma_{e/o} L}{2}-\omega t_0  \Big)\\
&\rightarrow -\beta_{e/o} V_0 L (1+\underset{\leftarrow}{\epsilon})\text{sin}\Big(\frac{k N^{EOM}_{e/o} L}{2}+\omega t_0  \Big)\\
\end{split}
\end{align}
\label{eomR}

Where we use the substitutions $\Gamma_{e/o}\equiv N_m-N^{EOM}_{e/o}$ and let $\Sigma_{e/o}\equiv N_m+N^{EOM}_{e/o}$ for here and below.  In the last two lines, terms higher than first order in the Taylor expansion of $\sin$ are kept in $\underset{\rightarrow}{\epsilon}$ and, below, in $\underset{\leftarrow}{\epsilon}$.  In the last line, a constant phase factor, $\tfrac{k N_m L}{2}$ has been removed from consideration.  If the optical signal is traveling against the RF wave, then $k N^{EOM}_{e/o} (L-z)=\omega (t-t_0)$ then the corresponding voltage equation and its integral is
\begin{align}
\begin{split}
\Delta\phi_{e/o} &(z)=\beta_{e/o} V_0 \sin(k (N_m+N^{EOM}_{e/o}) z -k N^{EOM}_{e/o} L-\omega t_0)\\
\phi^\leftarrow_{e/o}&=\int_0 ^L \Delta\phi_{e/o}(z) dz\\
&=\beta_{e/o} 2\frac{V_0}{k \Sigma_{e/o}} \text{sin}\Big(\frac{k\Sigma_{e/o} L}{2}\Big) \text{sin}\Big(\frac{k\Gamma_{e/o} L}{2} -\omega t_0\Big) \\
&= \beta_{e/o} V_0 L (1+\underset{\leftarrow}{\epsilon})\text{sin}\Big(\frac{k\Gamma_{e/o} L}{2} -\omega t_0\Big) \\
&\rightarrow -\beta_{e/o} V_0 L(1+\underset{\leftarrow}{\epsilon}) \text{sin}\Big(\frac{k N^{EOM}_{e/o} L}{2} +\omega t_0\Big) \\
\end{split}
\end{align}
\label{eomL}

Here, from the specification sheet for the EOM, $\Gamma_{e/o} \approx 10\% N_{e/o}\approx.2$ with $N^{EOM}_e=2.17$ and $N^{EOM}_o=2.25$.  $L=7.1 cm$ so at 5MHz drive frequency, $\tfrac{k \Gamma_{e/o} L}{2}\approx 8\times 10^{-4}$.  Since EOM drive voltage is adjusted such that optimal phase modulation is $.9$ rad, and because the $r_{33}$ and $r_{13}$ electrooptic coefficients, describing the modulation of the TM and TE modes of the EOM, are of the same order of magnitude, $\beta_{e/o} V_0 L\approx 1$ and the cubic error term to the Taylor series approximation for the sin function above is on the order $10^{-8}$ less than that of the first order term.  Each of these expressions for the modulation describes waves in the voltage signal traveling in a single direction for each term.  To include possible effects from a reflected RF signal as created by the known impedance mismatch in the modulator, then the resulting directionally dependent modulation amplitudes will be linear combinations of the two terms above, and this will only effectively change the amplitudes and phases for the modulation terms in the analogous form for Equation \ref{intensity}:

\begin{equation}
I(t)=\tfrac{1}{2}\mathcal{I} \lvert e^{i\phi_{m1}\text{sin}(\omega t+\vartheta_1) +\theta_K} \pm e^{i\phi_{m2}\text{sin}[\omega (t+\tau_{path})+\vartheta_2] -\theta_K} \rvert^2
\end{equation}

Expanding this expression can not lead to spurious terms leading to an anomalous Kerr signal.  However, the directional dependence for the phase modulations, implies, as previously mentioned, that there may be a corresponding directional dependence in the amplitude modulation and a spurious signal, and in this way, RAM will be introduced.

Estimating the resulting size of RAM can not be accomplished without a model for a specific mechanism that generates it.  One possible mechanism is amplitude modulation occurring from modulated scattering at the surfaces of the Lithium Niobate crystal.  The contributions to the amplitude modulation when the beam passes through the two facets of the EOM twice are given by:

\begin{align*}
A_1(t_0)&= (1+\alpha\text{sin}(\omega t))\\
&\quad\times(1+\alpha\text{sin}(\omega t+\phi_e))\\
&\quad\times(1+\alpha\text{sin}(\omega (t+\tau_{\text{path}})+\phi_e))\\
&\quad\times(1+\alpha\text{sin}(\omega (t+\tau_{\text{path}})+\phi_e+\phi_o))\\
A_2(t_0)&= (1+\alpha\text{sin}(\omega t))\\
&\quad\times(1+\alpha\text{sin}(\omega t+\phi_o))\\
&\quad\times(1+\alpha\text{sin}(\omega (t+\tau_{\text{path}})+\phi_o))\\
&\quad\times(1+\alpha\text{sin}(\omega (t+\tau_{\text{path}})+\phi_e+\phi_o))
\end{align*}

From Equations \ref{eomR} and \ref{eomL}, the optimal choice of frequency is $\pi=\omega\tau_{\text{path}}+\tfrac{1}{2}\phi_e+\tfrac{1}{2}\phi_o$.  As $\phi_{e/o}=N_{e/o} k L\approx 3\times 10^{-2}$ rad, expanding the above expressions will result in all terms containing first harmonics canceling out or contributing negligibly .

\subsection{RAM and Birefringence}

If light is scattered at the sample by birefringence into the wrong modes as in Figure \ref{Paths}c, then the two will not interfere and their power signals will just add at the detector.  Before, it was shown that the effects of RAM on the primary modes will cancel when the modulation signal is delayed by $\pi$ rad on the second pass through the EOM, as compared to the first pass.  When aligned in this way, this condition will not be the case for modes traveling along the incorrect paths, because the path lengths for these two modes will not be equal to that of the primary modes.  If we consider that some fraction of the light, $\rho$, is scattered at the sample into the incorrect paths, then we can estimate the size of the resulting spurious signals for RAM, $A( t)=1+\alpha\sin(\omega t)$, applied to the signal on each pass through the EOM.  As measured before, $\alpha=10^{-5}$ and the phase difference between the two beams of light after reaching the sample is given by $\epsilon=\omega L\tfrac{n_e-n_o}{c}=\pi\frac{\Delta \ell}{L}\sim 10^{-3}$.  Because of the incoherence of these two beams, the phase modulation does not need to be considered in the model below, so only amplitude modulation from the first and second passes through the EOM for both beams is included:

\begin{align*}
I(t)&=\rho\mathcal{I}\times\big(A(t)A(t+\tfrac{2Ln_e}{c})+A(t)A(t+\tfrac{2Ln_o}{c})\big)\\
&=\rho\mathcal{I}\times\big(1+\alpha\sin(\omega t))(2-\alpha \sin(\omega t+\epsilon)\\
&\hspace{1.4cm}-\alpha \sin(\omega t-\epsilon)\big)\\
&\approx 2\rho\mathcal{I}\times\big(1-\alpha^2\sin^2(\omega t)\big)+\rho\mathcal{I}\alpha\epsilon^2\sin(\omega t)
\end{align*}

Let $\rho=10^{-3}$, as found before by measuring the reduction in the second harmonic after removing the quarter-wave plate.  The expected contribution to the first harmonics is then down by a negligible factor of $10^{-14}$, demonstrating that this model can not account for the spurious signals.  However, in systems where the amplitude modulation and birefringence is greater, this effect might be relevant.

\section{Nonlinear Effects}
\label{nonlinearAppendix}

The nonlinear Kerr effect within the fiber or other components can be a source of nonreciprocity if the power in the two counter-propagating channels is different or the source is pulsed\cite{lefevre}.  Unlike RAM, because nonlinear effects will generate a phase difference within the two beams, the resulting Kerr signals will change their sign if the quarter-wave plate is rotated $90^\circ$.  Nonlinear effects within the fiber should not contribute because the nonlinear Kerr effect is coherent process.  For each polarization channel within the fiber, the two counter-propagating and broad-bandwidth beams will not be coherent and therefore the nonlinear Kerr effect will average to zero.  The only place where the nonlinearities may contribute to a spurious signal is where the beams are coherent, such as in the EOM.
As mentioned before, an intensity dependent Kerr signal is an indicator that a nonlinear effect is contributing to the measurement.  The intensity should be tuned in such a way as to avoid changing the spectrum of the SLED, which could potentially affect size of the offsets.  We have not seen clear evidence for this effect, thus suggesting that nonlinear responses in the instrumentation are not significant with power levels used in this experiment.  

Nonlinear optical effects can potentially induce magnetism in the sample: if such a sample is radiated with unequal power from $+$ and $-$ circularly polarized light will magnetize the sample, however the measured intensity difference between the two beams is low.  Another possibility is if the incident light significantly changes the sample's temperature or changes some other macroscopic property.  


\section{Alignment Procedure}
\label{alignment}
Aligning the all-fiber SI is performed by cycling through a series of steps until satisfactory performance is achieved.  

\begin{enumerate}[1)]
\item Fix the optical wavelength of the source.  Small changes in the center wavelength may affect the performance, however we find that the system is unaffected by shifts of $\pm$5nm.  It is also important that the spectral width of the source be broad: we have found that narrow band sources with spectral width less than 1 nm will introduce noise of order at least 10 $\mu$rad.  It is also essential that the optical components be designed such they are optimally functioning within the selected wavelength range.
\item Assemble the components and ensure that there is sufficient throughput to the detector.  Use a mirror in lieu of a magnetic sample and carefully ensure there are no stray magnetic fields present.
\item Spurious first harmonics can enter the system in the form of electronic noise coupling to the detector and the light source, so it is prudent to characterize these signals and eliminate them.  One method of distinguishing if the signal is originating from either the SLED or the detector is to measure the size of a spurious harmonic when illuminating the detector with the SLED and comparing it to the size of the signal when illuminating with a small battery-powered light source, such as a laser pointer.  
\item Set the modulation voltage of the EOM to give a $.92$ rad phase modulation depth by estimating it from the factory spec sheet.  Estimate and set the frequency which will give a $\pi$ phase shift given the length and index of refraction of the fiber.
\item Still using a mirror as a sample: Maximize the second harmonics by rotating the quarter-wave.
\item From Equation \ref{intensity}, the optimal value of $\omega$ is found by adjusting the EOM's driving frequency until the second harmonics are maximized.  Alternatively, as also prescribed by Equation \ref{intensity}, the second harmonics should vanish at the twice the optimal frequency.  The optimal frequency should ideally coincide with a minimum signal from spurious first harmonics, but depending on the noise source, this is not always the case.
\item On a homogeneous, strongly magnetic sample (i.e. CoPd multilayer film), maximize the first harmonic magnitude by tuning the modulation depth.  This guarantees that the value of the modulation depth is the half the value of the argument of the $J_1$ Bessel function that gives the maximum, 0.92.  This condition also maximizes the size of the first harmonics, making it easier to measure.
\item On a homogeneous, strongly magnetic sample, align the lock-in phase offset so that the first harmonic signal is entirely in phase with the reference.  When returning to a mirror, there should be no signal on the quadrature component of the first harmonics.  Note the phase of any spurious signals (i.e. from RAM) that are present after this alignment.
\item The RF signals on the co-axial cables between the electronics have transmission amplitudes and phases that are strongly dependent on frequency and cable length.  Once the driving frequency for the EOM has been selected, calibrate the transmission ratios for both harmonics going from the detector output to the lock-in amplifiers using a second RF source in place of the detector and phase-locked to the main function generator.
\item Determine which of the two orientations the wave plate is in by measuring the Kerr angle with a sample of known magnetization.
\item Check if RAM is causing offsets by examining how the Kerr signal varies as the quarter-wave plate is rotated. 
\item Test the system with birefringent samples and chiral samples and demonstrate that the measurement of the Kerr signal is immune to such perturbations.  Misalign and then realign the quarter-wave plate to demonstrate immunity to this, as well.  Measure the Verdet constant of a known sample to check the calibration.
\item Adjust the fiber layout until any spurious signals from RAM are minimized.
\end{enumerate}

Pay particular attention to the following issues:

\begin{enumerate}[A)]
\item The alignment of the miniature lens/quarter-wave plate assembly at the termination of the fiber at the sample is difficult.  We recommend carefully modeling the propagation of the beam through this optic using ray tracing software.  It is often helpful to diagnose problems and check components in the system using a fiber-coupled collimator, a free-space quarter-wave plate mounted on a rotation stage, and a mirror placed at the focal point of the beam.
\item Because the fiber only supports a single Gaussian mode per polarization per direction, it is essential that the wavefront is specularly reflected from the sample.  Any roughness or curvature on the surface will scatter light out of those optical modes that are matched to the fiber, substantially decreasing the amount of power returning to the detector.
\item The Kerr angle should be unaffected by changes in the total power through the whole system (as might occur when  samples of different reflectivities are measured).  This would otherwise indicate the presence of a spurious signal superposed on the desired one, but it could also indicate the presence of an intensity independent spurious harmonic from electronic noise as was described previously.  
\item  When dividing the first harmonic signal by the second harmonic signal it is important to ensure that digital quantization errors are not large which would otherwise result in this observation.  We avoid this by operating the lock-in which measures first harmonics in ``ratio" mode which normalizes its measurement by the output of the lock-in which measures the second harmonics.
\item The Kerr signal should be drift-free.  Plenty of optical isolation in front of the SLED will prevent back-reflections possessing RAM from being re-amplified.  Drift from thermal fluctuations within the EOM can also appear when the its drive frequency is not optimal for the given optical path length.
\item The first harmonic quadrature should also be close to zero; it, too, will likely not be exactly centered at zero, because of the variety of spurious offsets.  It is often helpful to create a real-time phasor plot of the first harmonics.  Once the alignment procedure is finished, the signal on the quadrature should be drift free.
\item The effects of thermal fluctuations can be examined by several tricks; touching the fiber, uncoiling it, or carefully heating it with a hair dryer.  These tests can be performed for the EOM as well, as it can also be sensitive to temperature changes.  We house the all the components in Styrofoam casings to minimize thermal expansion of the components and eliminate air currents.  We have noticed that thermal drifts from the EOM do seem to cause slow fluctuations in the Kerr signal on the scale of a 0.1 $\mu$rad, but they are improved by thermally anchoring the modulator housing.
\item Electronic noise from stray RF fields and ground loops can cause spurious first harmonics and its presence should be addressed using the same methods for dealing with ground loops\cite{Ott} (i.e. filters, universal power supplies, deadmen and re-arranging the positions of the electronic components and by trying different power outlets.  In particular, stray signals from the function generator can leak onto the SLED, the detector and cables running to and from the Lock-In Amplifiers.  Adding low pass filters on the outputs of the function generator or the inputs of the lock-in amplifiers can help.  Make sure that the enclosures for filters and power splitters between the detector, the function generator and the lock-in amplifiers are mounted in a fashion that leaves them ungrounded.  Poor cables and loose connections are especially susceptible to RF pick-up.
\end{enumerate}

\bibliographystyle{plain}

\end{document}